\documentclass[12pt,draft,onecolumn]{IEEEtran}

\usepackage{amssymb}
\usepackage{amsmath}
\usepackage{amsthm}
\usepackage{epsfig}
\usepackage{mathrsfs}

\newcommand{\R}{\mathbb{R}}    

\newcommand{\la}{\langle}
\newcommand{\ra}{\rangle}

\newcommand{\be}{\begin{equation}}
\newcommand{\ee}{\end{equation}}
\newcommand{\ba}{\begin{align}}
\newcommand{\ea}{\end{align}}
\newcommand{\bea}{\begin{eqnarray*}}
\newcommand{\eea}{\end{eqnarray*}}
\newtheorem{thm}{Theorem}[section]

\newtheorem{rem}[thm]{Remark}

\begin{document}
\title{A new method on deterministic construction of the measurement
matrix in compressed sensing}
\author{
Qun Mo
\thanks
{
Research supported in part by the NSF of China under grant
10971189 and 11271010, and by the fundamental research funds
for the Central Universities.
}
\thanks
{
Q. Mo is with the Department of Mathematics, Zhejiang University,
Hangzhou, 310027, China (e-mail: moqun@zju.edu.cn ).
}
}

\maketitle

\begin{abstract}
Construction on the measurement matrix $A$ is a central problem in compressed
sensing. Although using random matrices is proven optimal and successful in both
theory and applications. A deterministic construction on the measurement matrix
is still very important and interesting. In fact, it is still an open problem proposed
by T. Tao. In this paper, we shall provide a new deterministic construction method
and prove it is optimal with regard to the mutual incoherence.
\end{abstract}

\begin{IEEEkeywords}
Compressed sensing, measurement matrix, deterministic construction,
mutual incoherence, sparse signal reconstruction.
\end{IEEEkeywords}

\IEEEpeerreviewmaketitle

\section{Introduction}
\IEEEPARstart{S}{parsity} and compressed sensing have attracted a great deal of
attentions recently. The key idea in compressed sensing \cite{CRT,D} is that if a signal
$x \in \R^N$ is sparse, then we can exactly recover it from much fewer
measurements $b = Ax$, where $A \in \R^{m \times N}$ is the measurement
matrix and usually $m \ll N$.

To be more precise, we say $x \in \R^N$
is $s$-sparse if $\|x\|_0 \leq s$, where $\|x\|_0$ is the number of nonzero entries
of $x$. Also, we say $x$ is sparse if $x$ is $s$-sparse and $s \ll N$.
In many applications like image processing, video processing etc, signals are often
in a very high dimensional space, i.e., $x \in \R^N$ with a very large $N$. That is,
a signal $x$ usually has a huge mount of entries unknown. It would take lots of
effort to measure these entries if we measure them one by one. Fortunately, due to
their natural structure, many signals are sparse
or can be well approximated by sparse signals, either under the canonical basis
or other special basis/frames.

For simplicity, we assume that $x$ is sparse under the canonical basis in $\R^N$,
that is, $x = \sum_{k = 1}^s x_{j_k} e_{j_k}$ with
$1 \leqslant j_1 < j_2 < \cdots < j_s \leqslant N$. An important remark is that usually
we do not have any prior information or assumption about the exact location of these
nonzero entries of $x$.

To retrieve such a sparse signal $x$, a natural method is to solve the
following $l_0$ problem
 \be\label{p0}
 \min_{x} \|x\|_0 \quad \text{subject to} \quad Ax = b
 \ee
where $A$ and $b$ are known. To ensure the $s$--sparse solution is
unique, we would like to use the \emph{restricted isometry
property} (RIP) which was introduced by Cand\`es and Tao in
\cite{CT}. A matrix $A$ satisfies the RIP of order $s$ with the
\emph{restricted isometry constant} (RIC) $\delta_s = \delta_s(A)$
if $\delta_s$ is the smallest constant such that
 \be\label{RIP}
 (1-\delta_s)\|x\|_2^2
 \leq
 \|A x \|_2^2
 \leq
 (1+\delta_s)\|x\|_2^2
 \ee
holds for all $s$-sparse signal $x$.

If $\delta_{2s}(A) < 1$, the $l_0$ problem has a unique $s$-sparse
solution \cite{CT}. The $l_0$
problem is equivalent to the $l_1$ minimization problem when
$\delta_{2s}(A) < \sqrt{2}/2$,
please see \cite{C, ML, CZ} and the references therein.

Now it is natural to ask how to construct a desired measurement matrix.
Using random matrix is proven to be very successful. Candes and Tao have
proven the following theorem:

\begin{thm}
If the elements of a matrix $A$ is independently drawn from the gaussian distribution
${\cal N}(0, m/N)$, then with very high probability, we have
$\delta_{2s}(A) \leqslant m/(s \log N)$.
\end{thm}

A further conclusion of the above theorem is: If the elements of $A$ is
independently drawn from the gaussian distribution ${\cal N}(0, m/N)$ and
$m \geqslant C s \log N$ with some constant $C$, then with very high probability,
we have $\delta_{2s}(A) < \sqrt{2}/2$.

Since reducing the number of measurements is essential in compressed sensing.
It is highly desirable to construct a measurement matrix $A \in \R^{m \times N}$
with $m$ as small as possible while satisfying $\delta_{2s}(A) < \sqrt{2}/2$.
From the other hand, it is also highly desirable to construct a measurement matrix
$A$ with optimal instances. It is proven \cite{CDV} that to satisfy optimal instance,
we must have $m \geqslant C s \log N$ with some constant $C$. Therefore, using
random  matrix to construct the measurement matrix $A \in \R^{m \times N}$ with
$m = C s \log N$ is optimal with regard to optimal instances.

Although using random matrix is so successful, it is still very important and
interesting to study deterministic constructions. In fact, it is still an open problem
proposed by Tao. 

\section{Mutual incoherence}

As mentioned before, $\delta_{2s} < 2^{-1/2}$ is a sharp sufficient condition.
However, according to its definition, R.I.C. is very hard to calculate. On the other
hand, another constant, mutual incoherence, is much easier to calculate. For a
measurement matrix $A$, we denote $\mu_{A}$ the mutual incoherence by
\be
\mu_A := \max_{1 \leqslant i < j \leqslant N}
    \dfrac{|\la A e_i, A e_j \ra}{\| A e_i \|_2 \| A e_j \|_2}
.
\ee
It is proven by Tai and Wang [***] that if
\be
\label{MIPcond}
\mu_A < 1/(2s - 1),
\ee
then the measurement matrix $A$ is suitable for recovering every $s$-sparse signal
$x$ from $b = Ax$ and the above condition is sharp.
Now we will focus on how to construct a sensing matrix $A \in \R^{m \times N}$
such that \eqref{MIPcond} is satisfied.

First of all, let us review the possible range of $m$ when $N$ and $s$ are given.
Define
\be
\label{MIPdef2}
\mu_{m, N} := min_{ A \in \R^{m \times N} } \mu_A
.
\ee,
By the famous Welch bound [***], we have
\be
\label{Welch_bound}
\mu_{m, N} \geqslant \sqrt{\dfrac{N - m}{(N - 1)m}}
.
\ee
However, it is not a sharp bound in some situations. For instance, if we
fix $m$, then all the column vectors of $A$ are in $\R^m$. Now when
$N \rightarrow +\infty$ , which means the number of column vectors
of $A$ goes to positive infinity, then the mutual incoherence of $A$ will go
to 1, since those column vectors of $A$ are getting crowder
and crowder in $\R^m$. That is,
$$
\lim_{N \rightarrow +\infty} \mu_{m, N} = 1
$$
when $m$ is fixed.
If \eqref{Welch_bound} is sharp, we would have
$$
\lim_{N \rightarrow +\infty} \mu_{m, N} = \sqrt{\dfrac{1}{m}}
$$
which contradicts the above equality!
As pointed out by [****], another bound is
\be
\label{width_bound}
m \geqslant C \ln N (\dfrac{1}{\mu})^2 / \ln (\dfrac{1}{\mu})
\ee
where
$C$ is a constant independent of $m$, $N$ and $\mu$.
A remark of above bound is that it implies $\mu \rightarrow 1$
when $m$ is fixed and $N \rightarrow +\infty$.
Also, it is still unknown whether the above bound is sharp.
For all known constructions in the literature, we have
\be
\sqrt{m} \ln m \geqslant C \ln N \dfrac{1}{\mu}
\ee

Now we propose the following new construction method. This algorithm
is a random algorithm, the possibility of this algorithm to succeed is very
high. Moreover, if this random algorithm succeeds, we know for sure
that the output matrix $A$ will satisfy
the condition $\mu_A \leqslant \dfrac{1}{2s}$.

Algorithm:
1. Input $N$ and $s$. Fix the seed of a random generator.
Choose $m \geqslant \lceil{8s^2 \ln (2sN/\pi)}\rceil + 2$
and define $j = 0$, $x_0 = e_1 \in \R^m$.

2. Repeat the following:

2.1 Let $k := 0$ and replace $j$ by $j + 1$.

2.2 Replace $k$ by $k + 1$. Use the random generator to get a unit vector
$y \in \R^m$,
then calculate $\mu_y := max_{1 \leqslant i \leqslant j} | \la x_i, y \ra |$.

2.3 repeat 2.2 if $\mu_y > \dfrac{1}{2s}$ and $k < 10$.

3. Repeat 2 if $k < 10$ and $j < N$.

Now we claim the following theorem:

\begin{thm}
The possibility that the above algorithm find the desired sensing matrix
is at least $1-10^{-4}$. If the algorithm succeed, the worst computational
complexity is $10m N + N(N - 1)/2$.
\end{thm}

Proof:
For a given unit vector $x \in \R^m$, consider these two caps
$$
C_{x, s} := \{y \in \R^m | \| y \|_2 = 1 \hbox{ and } | \la y, x \ra |
\geqslant  1/(2s) \}
.
$$
By direct calculation, one can verify that
the surface area $A_{s, m} $ of these two caps $C_{x, s}$ satisfies
$$
A_{s, m} \leqslant 2s ( 1 - 1/(2s)^2 )^{(m - 1)/2} V_{m - 1}.
$$
Therefore, the possibility of finding a wanted $y$ such that
$$
| \la y, x_i \ra | \leqslant 1/(2s) \;\; \forall i = \{1, 2, \cdots, j \}
$$
is at least $1 - j A_{s, m}/A_m
\geqslant 1 - 2j s ( 1 - 1/(2s)^2 )^{(m - 1)/2} V_{m - 1}/A_m
\geqslant 1 - 2j s ( 1 - 1/(2s)^2 )^{(m - 1)/2} / (2 \pi)$,
which by direct calculation, is at least
$1 - ( j/(2N) )^{10}$.
Then the all claim can be verified by direct calculation.

\begin{rem}
It is only a very rough draft, a refined version with suitable
citations will be updated soon.
\end{rem}

\begin{biographynophoto}
{Qun Mo} was born in 1971 in China. He has obtained a B.Sc. degree in 1994
from Tsinghua University, a M.Sc. degree in 1997 from Chinese Academy of
Sciences and a Ph.D. degree in 2003 from University of Alberta in Canada.

He is current an associate professor in mathematics in Zhejiang University.
His research interests include compressed sensing, wavelets and their applications.
\end{biographynophoto}

\end{document}